\documentclass[doublecol]{epl2}
\usepackage{graphicx}
\usepackage{amsmath}
\usepackage{amsfonts}
\usepackage{amssymb}
\usepackage{epsf}
\usepackage{hyperref}
\usepackage{pstricks,pst-coil,pst-fill,pst-plot}

\begin{document}

\title{Probing surface states with many-body wave packet scattering}

\author{F. Damon\inst{1}, B. Georgeot\inst{1} \and D. Gu\'ery-Odelin\inst{2}}
\shortauthor{F. Damon, B. Georgeot and  D. Gu\'ery-Odelin} 
\institute{
 \inst{1} Laboratoire de Physique Th\'eorique, IRSAMC, Universit\'e de Toulouse, CNRS, UPS, France\\
\inst{2} Laboratoire Collisions, Agr\'egats, R\'eactivit\'e, IRSAMC, Universit\'e de Toulouse, CNRS, UPS, France
}
 \date{\today}

\abstract{The scattering of 1D matter wave bright solitons on attractive potentials enables one to populate bound states, a feature impossible with noninteracting wave packets. Compared to noninteracting states, the populated states are renormalized by the attractive interactions between atoms and keep the same topology. This renormalization can even transform a virtual state into a bound state. By switching off adiabatically the interactions, the trapped wave packets converge towards the true noninteracting bound states. Our numerical studies show how such scattering experiments can reveal and characterize the surface states of a periodic structure whose translational invariance has been broken. We provide evidence that the corresponding 3D regime should be accessible with current
techniques.} 
\date{December 21, 2015}

\maketitle

\section{Introduction}

The development of cold atom physics has led to many new advances in the recent past. 
In particular, the new field of atom optics using propagative matter waves achieved impressive advances, enabling to realize with cold atoms many features previously developed in optics, such as quantum reflection, beam splitters or Bragg reflection on optical lattices \cite{bragg}.

However, an important difference with optics is that cold atoms can be prepared in regimes where interaction is important. In this case, the physics of the system in the mean field approximation is described by the Gross-Pitaevskii equation \cite{pitaev,livredgo} (a similar equation also appears in nonlinear optics \cite{nonlinoptics}). This equation is nonlinear contrary to the usual Schr\"odinger equation. As such, it possesses for attractive interactions special types of solutions called solitons. In contrast with wave packets of the usual linear Schr\"odinger equation which spread with time, the variance of solitons in free space remains constant. Solitons are known in classical physics in many contexts, especially fluid mechanics \cite{fluids,refsol,Malomed89}. In the field of dilute Bose-Einstein condensates, they have been predicted theoretically \cite{refsol, Malomed89,pikovsky} and observed experimentally with rubidium-85 and lithium-7 atoms in various forms: dark \cite{darksolitonsexp}, bright \cite{brightsolitonsexp} and band-gap \cite{bandsolitonsexp} solitons.

In the present paper, we wish to put forward a useful aspect of these many-body wave functions.  Indeed, the scattering of solitons can be used to probe virtual and/or bound states in various potentials through scattering experiments, revealing structures which cannot be observed with an interaction-free wave packet. Indeed, in many-body wave packets part of the energy corresponds to the interaction energy and this creates a new freedom which can be used to populate such states.  It has been shown already theoretically that soliton scattering on some potentials was able to detect and characterize the energy of some bound states associated to localized defects \cite{Stoychev}, square potentials \cite{ernst}, impurity modes \cite{Izuka,Brazhnyi} and lattice defects \cite{Goodman}. Solitons were also studied as tools for interferometry \cite{Robins}. 
Some recent experiments have started the investigation of the scattering of solitons on short-size potential wells  \cite{cornish}.

In finite-size optical lattices, which are commonly constructed in cold atom experiments, the presence of boundaries at the edge of the potential breaks the translational invariance and results in the existence of surface states. They have been characterized and studied in condensed matter since a long time \cite{shockley,tamm}, and are of great importance in several areas such as semiconductor physics. In this paper, we will show how to observe and characterize such surface states through the scattering of solitons. 

First, we present the scattering of a soliton on a single well. We then extend the results on a collection of identical wells corresponding to a finite-size lattice potential which can be easily created by interfering laser beams in cold atom experiments, and discuss the observation of surface states.

\section{Solitons}

We consider hereafter exclusively bright solitons described by the following 1D Gross-Pitaevskii equation for the wave function $\psi(x,t)$:

\begin{equation}
i\hbar\frac{\partial}{\partial t}\psi=\left(-\frac{\hbar^2}{2m}\frac{\partial^2}{\partial x^2}+V(x)+g_{1D}N|\psi|^2\right)\psi\label{gp1d}
\end{equation}

Interactions between atoms are accounted for by the interaction strength $g_{1D}=2a\hbar\omega$, $N$ is the number of atoms, $a<0$ the scattering length, and $\omega$ accounts for the transverse trap frequency in the 3D to 1D reduction of dimensionality where $a$ is the 3D scattering length. In a purely 1D view, $g_{1D}$ has to keep the homogeneity of an energy times a length. 

This nonlinear equation admits stable solutions in free space called solitons, of the form (for a soliton started at $t=0$ and $x=0$ with mean velocity $\bar v$): 

\begin{equation}
\psi(x,t)=\sqrt{\frac{N|a|}{2\sigma^2}}\frac{\exp\left[i\displaystyle \frac{m\bar{v}}{2\hbar}(2x+(\bar{v}-Na\omega)t)\right]}{\cosh\left(\displaystyle \frac{N|a|}{\sigma^2}(x-\bar{v}t)\right)},\label{soliton}
\end{equation}
with $\sigma=\sqrt{\hbar/m\omega}$. The variance of the soliton is
 independent of time \cite{Malomed89}, contrary to the case of a noninteracting wave packets.  




\section{Single potential well}

As a first example, we study in the following the scattering of a soliton on a single well. We choose as potential the restriction of a sinusoidal optical lattice to a single period:
\begin{equation}
U(x)=-U_0\cos^2\left(\frac{\pi x}{d}\right)\quad\text{for}\quad -d/2\le x\le d/2,\label{pot-sin}
\end{equation}
whose first derivative is everywhere continuous. For numerical simulations, we use a typical experimental value $d=0.65 \mu m$  corresponding to possible experiments with rubidium 85. In the absence of interactions, the potential \eqref{pot-sin} exhibits bound states whose number depends on the relative depth $U_0/E_R$ with $E_R=\hbar^2k_R^2/(2m)$ and $k_R=2\pi/d$. For noninteracting wave packets, such bound states cannot be populated in a scattering experiment. Only indirect signatures of virtual or/and bound states exist e.g. scattering resonances and even zero-energy resonances \cite{dgo97}, or the transparency effect commonly referred to as the Ramsauer-Townsend effect.


\begin{figure}[]
\centering
\includegraphics[width=8cm]{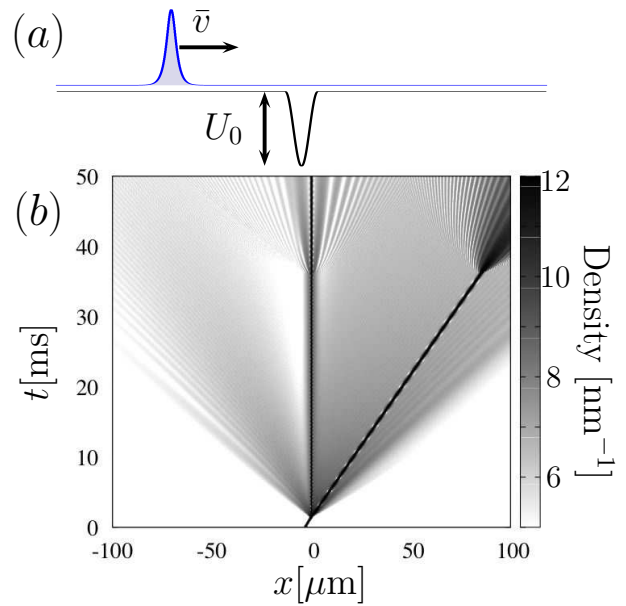}
\caption{(color online) (a) Scattering of a soliton (blue/gray curve) of mean velocity $\bar v$ onto the potential well (\ref{pot-sin}) (black). (b) Scattering of a soliton of $\bar v=0.3 v_{\rm R}$ (where $v_{\rm R}= \hbar k_R/m$) and $N|a|=4.5 \times 10^{-5}$ m, onto potential (\ref{pot-sin}) of depth $U_0=0.5E_{\rm R}$. The density is plotted as a function of time in log scale to enhance the contrast. Interactions are adiabatically switched off from $t=36$ ms (see text).}
\label{fig1}
\end{figure} 

Scattering of an interacting wave packet on such a potential cannot be described as usual through asymptotic distributions at large distances since trapping is also present. This is illustrated by our numerical simulations presented in Fig.~\ref{fig1}, where a soliton of mean velocity $\bar v$ launched from the left splits into three parts in the course of the scattering. 
For $t< 36$ ms, the reflected part moves to the left and expands, indicating that the corresponding density has no solitonic character, the interaction energy being too small to compensate for the kinetic energy. Another part is transmitted and propagates without spreading, indicating that another soliton has been formed with a fraction of the original atoms.  At last, a significant fraction of the atoms remains trapped for a long time in the potential well.

\begin{figure}[t!]
\centering
\includegraphics[width=8cm]{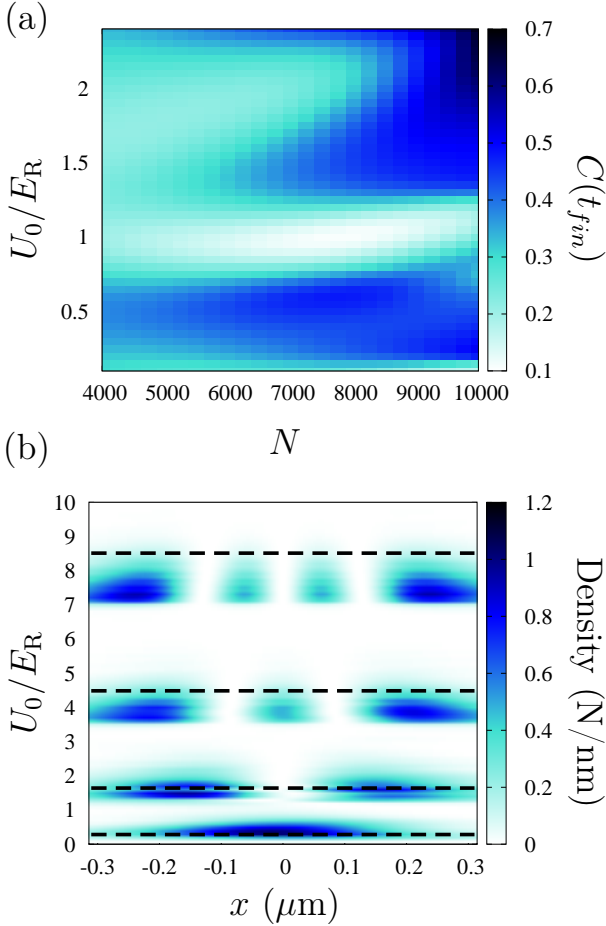}
\caption{(color online) (a) Proportion $C(t)$ of atoms in the potential well area for large $t$ values after scattering of a soliton of initial velocity $v=0.3 v_R$ and $N$ atoms as a function of the potential depth $U_0$. (b) Density plot for $N|a|=4.5 \times 10^{-5}$ m inside the potential well for long time (but before the switching off of the interaction). The black dashed lines represent the values for which a new bound state appears in the absence of interaction. Note that in the presence of interaction the populated bound states are below the interaction-free ones, but keep their topology.}
\label{fig1bis}
\end{figure} 

In order to separate these three parts, we consider the wave function probability of presence in three different zones:
$R(t)=\int_{-\infty}^{ -d/2} {\rm d} x|\psi(x)|^2 $, $C(t)=\int_{-d/2}^{d/2} {\rm d} x|\psi(x)|^2$
and $T(t)=\int_{d/2}^\infty {\rm d} x|\psi(x)|^2$.
In this way,  for sufficiently long times $R(t)$ will contain only the reflected part, $T(t)$ the transmitted part, and $C(t)$ the fraction of atoms which remains confined in the potential after the scattering.

Figure~\ref{fig1bis}(a) shows the trapped part of the probability for long time as a function of the number of atoms $N$ and the potential depth $U_0$. Depending on these parameters, this trapped part can vary over a large span of values, with thresholds that appear for certain critical values of the potential depth. 
The threshold values are related to the appearance of a bound states inside the well. The bound states keep the topology (number of nodes) of their interaction-free counterpart but are renormalized by the attractive interactions. The threshold values observed for populating renormalized bound states are thus below the values corresponding to the appearance of a bound state without interaction (see Fig.~\ref{fig1bis}(b)). A remarkable consequence of such a renormalization lies therefore in the possibility of populating virtual states whose energy are slightly positive and become negative as a result of the dressing of the state by the attractive interactions. 

\begin{figure}[t!]
\centering
\includegraphics[width=8cm]{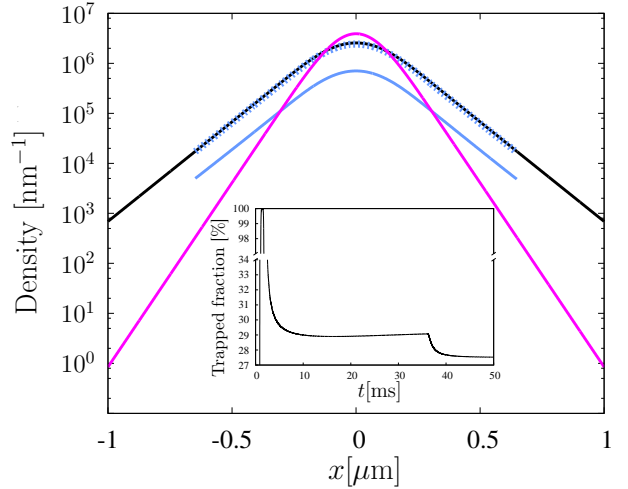}
\caption{(color online) Atomic density at $t=50$~ms inside the potential well (after the switching off of interactions) from the numerical dynamics (light blue/gray solid line), compared to the (normalized to one) square modulus of the fundamental eigenstate of the well with interaction and thus a smaller width (magenta/gray) and the one without interaction (black line). The dashed blue line which is perfectly superimposed to the black line corresponds to the numerical trapped distribution once renormalized to unity. The inset plot represents the variation of the part of the density inside the potential well $C(t)$ as a function of time. The switch off of the interactions yields a small decrease of the trapped fraction. Interacting bound states were calculated using the split operator method in imaginary time.}
\label{fig2}
\end{figure}

With the use of the Feshbach resonance such as in the case of lithium-7 atoms \cite{livredgo}, it is possible to cancel out the interactions.
We implement this process in our simulations by switching off the interactions in $\tau_{\rm switch}=1$ ms from $t=36$ ms. This duration has been chosen to ensure an adiabatic switching off of the interactions. Indeed, to populate a bound state in a 1D well of size $d$, the energy of the soliton should be on the order of the energy of the bound state $E_{\rm bound}\sim \hbar^2/2md^2$. Parameters have thus been chosen so that $\tau_{\rm switch} \gg \hbar / E_{\rm bound} \sim 30\;\mu$s. As a result, a fraction of the wave function leaves the potential well, but some atoms remain in the well for long time (see Fig.~\ref{fig1} and inset of Fig.~\ref{fig2}). This trapped part coincides with the noninteracting bound state of the potential well (Fig.~\ref{fig2}). For the specific example we are considering, it is possible to populate state by state each bound state. This occurs because the energy difference between two adjacent bound states remains large compared to the negative energy stored in the interactions.


\section{Finite-size lattice: surface states}


\begin{figure}[h!]
\centering
\includegraphics[width=8cm]{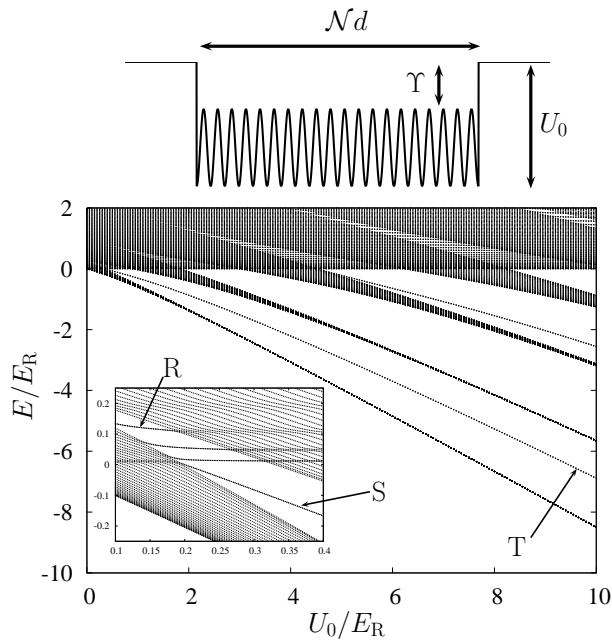}
\caption{(a) Schematic representation of the potential (\ref{potfini}). (b) Eigenvalues of the Hamiltonian for a single
particle in the potential (\ref{potfini}) as the function of the potential depth. (T) denotes a Tamm surface state, (R) a resonant surface state, and (S) a Schockley surface state.}
\label{fig3}
\end{figure}

In this section, we study the scattering of a soliton on a finite-size lattice. Such a lattice can be produced by interfering two mutually coherent laser beams. We first consider the following periodic potential with a square envelope (see Fig.~\ref{fig3}(a)):
\begin{eqnarray}
U(x)&=&-U_0\left[\sin^2\left(\frac{\pi x}{d}+\varphi\right)E(x)-\Upsilon \right] \label{potfini} \\
E(x)&=& \frac{1}{2}
\left[H\left(x\right)+H\left(\mathcal{N}d-x\right)\right], \nonumber
\end{eqnarray}
where $H(x)$ is the Heaviside step function, $\mathcal{N}$ the number of sites (wells of size $d$) and $\Upsilon$ is an additional offset term. 

In infinite periodic potentials, Bloch theory states that the eigenvalues are grouped into bands separated by gaps and correspond to extended states. However, for finite-size systems such as \eqref{potfini} additional states appear that are localized at the potential edges. 
Contrary to the generalization of Bloch states to such finite-size lattices, which are exponentially decreasing but only outside the lattice, the surface states decrease exponentially on both sides of the potential edge. In the limit of shallow potentials (small $U_0/E_R$) they have been characterized in \cite{shockley} and are called Shockley states, whereas in the limit of deep potentials (large $U_0/E_R$) they have been described in \cite{tamm} and are called Tamm states. Both states are bound states and appear inside the gaps separating the bands of Bloch states. At positive energy, other analogous states appear as resonances. These three kind of surface states are illustrated in Fig.~\ref{fig3}. 

\begin{figure}[t!]
\centering
\includegraphics[width=8cm]{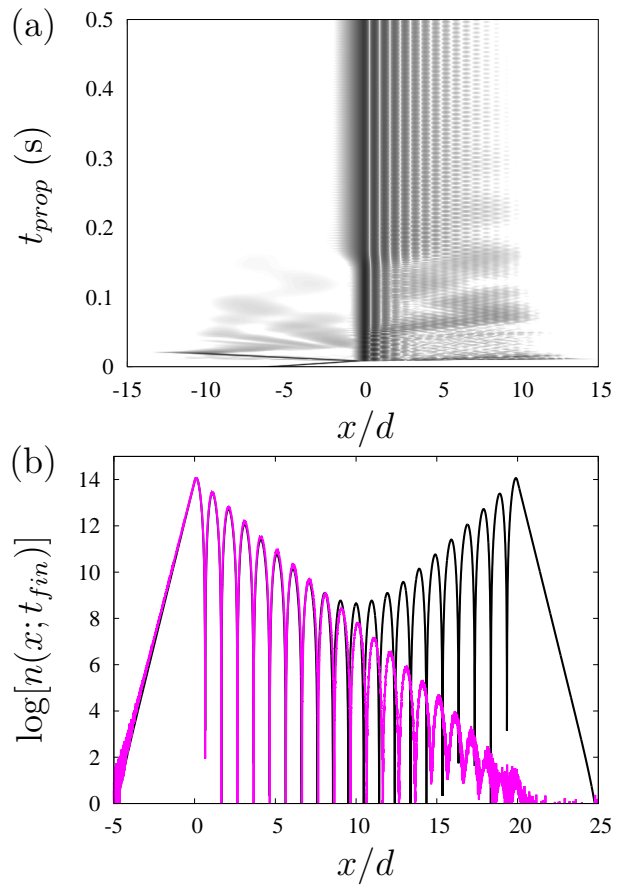}
\caption{(color online) (a) Scattering of a soliton of mean velocity $\bar v=0.1 v_{\rm R}$ with $N|a|=3. \times 10^{-5}$ m onto the potential (\ref{potfini}) of depth $U_0=0.5E_{\rm R}$. The density of gray represents the logarithm of density of atoms as a function of the position and time. Interactions are switched off adiabatically at $t_{prop}=150$~ms. (b) Final density distribution of atoms at $t_{fin}=1$~s (magenta/gray line) compared to the corresponding noninteracting surface state for the same potential (black curve).}
\label{fig4}
\end{figure}

\begin{figure}[h!]
\centering
\includegraphics[width=8cm]{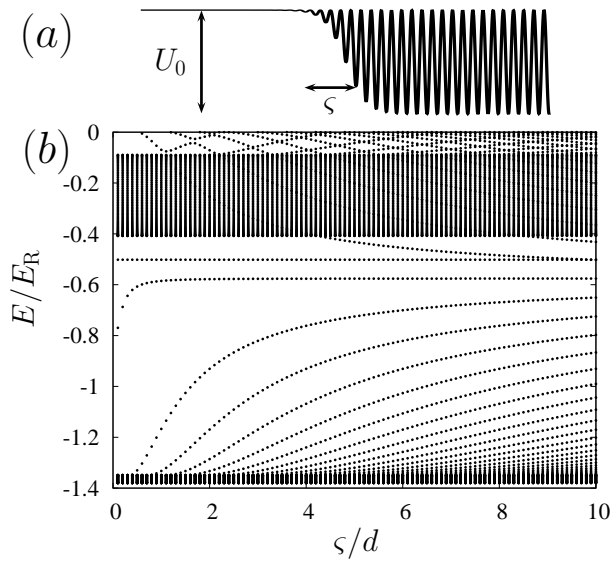}
\caption{(a) Schematic representation of the potential (\ref{slope}). (b) Energy levels distribution for this
potential as a function of the smoothing parameter $\varsigma$, for a depth of $2E_{\rm R}$. 
}
\label{fig5}
\end{figure}

In our numerical experiment, we launch a soliton on potential \eqref{potfini}. We observe a great variety of behaviors depending on the soliton total energy and the depth of the potential: the wave packet can be reflected, transmitted across the lattice, or trapped at the boundary. As previously, we relate these behaviors to the presence or absence of bound states which can trap the soliton at a given energy.
We thus expect that if a surface state is present the soliton can be trapped at the potential edge.

Such a phenomenon is visible in Fig.~\ref{fig4} (a). After scattering of the soliton on the finite-size lattice potential, part of the wave packet is reflected, part is transmitted, but a substantial fraction of the atoms remain trapped at the potential edge for long time. In order to confirm that the trapped part corresponds to a true surface state, we adiabatically switch off the interactions at $t=t_{prop}=150$ ms \footnote{The depth of the potential has also been lowered in order to induce a fast disappearance of the components of the wave packets that are contaminated by the extended states.}. The result is shown in Fig.~\ref{fig4} (b): the final wave function on the potential edge coincides exactly with the left half of a true surface state, with exponential decrease of the envelope on both sides of the potential edge.

The envelope of potential \eqref{potfini} has two discontinuities. To validate the experimental feasibility of our study, it is worth exploring how the preceding results are modified when the discontinuities are smoothened over a length $\varsigma$ (see Fig.~\ref{fig5} (a)):  
\begin{equation}
U(x)=-\frac{U_0}{2}\sin^2\left(\frac{2\pi x}{d}+\phi\right)\left[1-\tanh\left(\frac{x}{\varsigma}\right)\right].\label{slope}
\end{equation}

The corresponding energy spectrum is shown on Fig.~\ref{fig5} (b).  One recovers the results of the preceding paragraphs for $\varsigma \rightarrow 0$: a single bound surface state is visible in the gap. When  $\varsigma$ increases, this particular state remains unchanged, but more and more additional surface states leave the bands and go inside the gap. This confirms the existence of surface states for a periodic potential with a smooth envelope.

\begin{figure}[t!]
\centering
\includegraphics[width=7.cm]{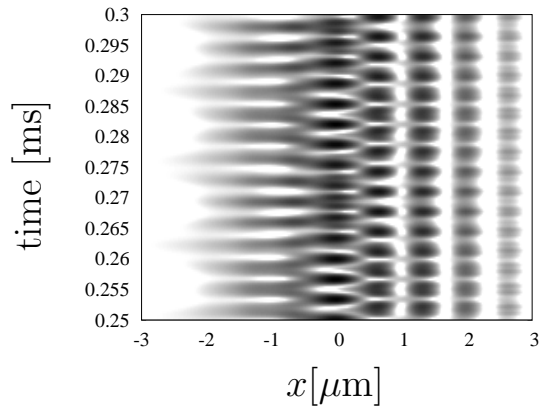}
\caption{Density at the potential edge of (\ref{slope}), where surface states are maximal and after the adiabatic switch off of interactions at $t=150$ ms. Parameters: soliton with $N|a|=3. \times 10^{-5}$ m, mean velocity $\bar v=0.27v_{\rm R}$, potential
(\ref{slope}) of depth $U_0=5E_{\rm R}$ and smoothing length $\varsigma=2d$ (the density is in logscale). 
}
\label{fig6}
\end{figure}

We have numerically simulated the scattering of a soliton on such a potential. Reflected and transmitted waves are still present and a substantial part of the atoms remains trapped. A closer look to the trapped part reveals the presence of a beating pattern (see Fig.~\ref{fig6}). Indeed, the presence of several surface states with small energy differences in the gap makes possible the trapping of the atoms in a superposition of surface states. The frequency of this beating pattern is directly related to the energy difference between the surface states populated, and allows to realize a direct spectroscopy of these states.  We note that the topology (number of nodes) of the states are preserved after free expansion and can be observed in time-of-flight experiments.

\section{Experimental feasability}

\begin{figure}[h!]
\centering
\includegraphics[width=8cm]{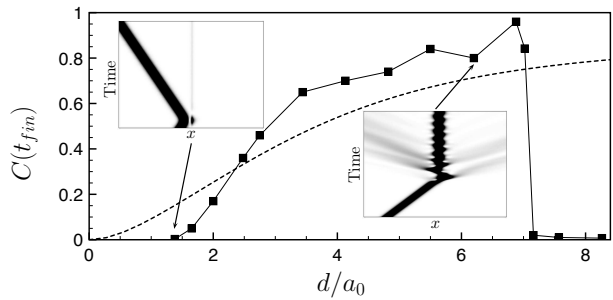}
\caption{Trapped population (black square) through the scattering of a soliton on the attractive potential (\ref{pot-sin}). Parameters: $\bar v=0.2$ mm/s, $a= -5$ nm, $\omega=2\pi \times 879$ s$^{-1}$ and $N=36$ atoms ($a_0=363$ nm). The dashed line results from a qualitative energetic argument (see text). Insets: snapshots of the probability density as a function of time for two different values of $d$: $d=1.4 a_0$ and $d=6.2a_0$.}
\label{figlast}
\end{figure}

The experimental implementation requires the transposition of the ideas presented here in a 3D situation.  Let us work out a simple energetic argument to provide a sufficient condition to populate a bound state (energy $E_{\rm bound}<0$). Consider an incoming soliton of mean velocity $\bar v$ and scattering length $a<0$, made of $N$ atoms with therefore an energy $E_I=Nm\bar v^2/2 + S_0 \omega^2N^3$ with $S_0=-ma^2/6$. We assume that after the scattering $(N-N^\prime)$ atoms populate the bound state: they have an energy $E_T=(N-N^\prime)E_{\rm bound} + S_0\omega^2_{\rm bound}(N-N^\prime)^3$, where $\omega_{\rm bound}$ characterizes the bound state. The $N^\prime$ other atoms are assumed to fly away from the scattering region with a velocity $\pm \bar v$ and an extra internal energy $\Delta E$. Their energy therefore reads $(N^\prime m \bar v^2/2 + \Delta E)$. By energy conservation, $\Delta E=E_I-E_T-N^\prime m \bar v^2/2$. For $\Delta E<0$, we shall assume that the soliton exists even in an excited state (see for instance the breathing mode in Fig.~\ref{fig1}).  This constitutes our simple condition for loading atoms in the bound state. We propose in the following to work out quantitatively a concrete example using the potential (\ref{pot-sin}) with a depth $U_0=E_R=h^2/2md^2$ and  a soliton width satisfying $N|a|=a_0/2$ i.e. the 3D stability criterion. Results are summarized in Fig.~\ref{figlast}. We observe a threshold value of $d$ above which the soliton starts to populate te trap. For $d<3 a_0$, the remaining part of the soliton is reflected (similarly to the quantum reflection phenomenon). For larger values a transmission is observed and an increasing number of atoms is trapped. The energetic argument described above is in rather good agreement with simulations. For large value of $d$, we expect that the soliton follows "adiabatically" the potential and does not populate the bound state. Simulations show that the transition between those two regimes occurs in a rather sharp manner. Those results are generic.

\section{Conclusion}

In this paper, we have shown that the scattering of solitonic wave packets built from interacting atoms can be used to probe and populate virtual and/or bound states of potentials, which could not be performed in the absence of interactions. 
We have shown that in the case of a single well potential, it is possible to detect the bound states of the potential and to trap a fraction of the atoms in the well for specific values of the potential depth. The threshold where the interacting atoms are trapped appears below the bound state energy for noninteracting atoms. The adiabatic switch off of the interaction enables one to make the density converge to the correct noninteracting bound state, starting from interacting atoms.

These results can be extended to the case of surface states appearing at the potential edge of a finite-size optical lattice. 
In the experimentally relevant case of a slowly varying enveloppe, where we have shown that many surface states are present, it is possible to trap the atoms on a superposition of surface states and use the beating pattern of the wave function to perform a spectroscopy of these states.

Our study has been carried out in the framework of the Gross-Pitaevskii equation. An extension to a complete quantum treatment is a priori necessary to confirm quantitatively our predictions \cite{castin2009,weiss2014,opticsexpress2012}. The scattering of solitons is a versatile tool to probe the structure of potentials; the imaging of the wave function allows to obtain spectroscopic information as well as density information, on both interacting or noninteracting bound states. These effects are at reach with current experimental technique.

We thank CalMiP for access to its supercomputers and the University Paul Sabatier (OMASYC project). We thank Yvan Castin for useful discussions. This work was supported by Programme Investissements d'Avenir under the program ANR-11-IDEX-0002-02, reference ANR-10-LABX-0037-NEXT.

\end{document}